\documentclass[conference]{IEEEtran}
\IEEEoverridecommandlockouts

\usepackage{cite}
\usepackage{amsmath,amssymb,amsfonts}
\usepackage{algorithmic}
\usepackage{graphicx}
\usepackage{textcomp}
\usepackage{xcolor}
\usepackage{placeins}
\usepackage{fixltx2e}
\usepackage{bbding}
\usepackage{multirow}

\usepackage[T1]{fontenc}
\usepackage{makecell}
\usepackage[utf8]{inputenc}
\usepackage{bm}

\def\BibTeX{{\rm B\kern-.05em{\sc i\kern-.025em b}\kern-.08em
    T\kern-.1667em\lower.7ex\hbox{E}\kern-.125emX}}
\begin{document}

\title{Prototype based Masked Audio Model for Self-Supervised Learning of Sound Event Detection}

\author{
    \IEEEauthorblockN{Pengfei Cai$^{1}$, Yan Song$^{1}$,  Nan Jiang$^{1}$, Qing Gu$^{1}$, Ian McLoughlin$^{2}$} 
    \IEEEauthorblockA{$^1$ National Engineering Research Center of Speech and Language Information Processing, University of \\ Science and Technology of China, China \quad $^2$ICT Cluster, Singapore Institute of Technology, Singapore }
    \IEEEauthorblockA{cqi525@mail.ustc.edu.cn, songy@ustc.edu.cn}
}

\maketitle

\begin{abstract}

A significant challenge in sound event detection~(SED) is the effective utilization of unlabeled data, given the limited availability of labeled data due to high annotation costs.
Semi-supervised algorithms rely on labeled data to learn from unlabeled data, and the performance is constrained by the quality and size of the former.
In this paper, we introduce the Prototype based Masked Audio Model~(PMAM) algorithm for self-supervised representation learning in SED, to better exploit unlabeled data.
Specifically, semantically rich frame-level pseudo labels are constructed from a Gaussian mixture model~(GMM) based prototypical distribution modeling.
These pseudo labels supervise the learning of a Transformer-based masked audio model, in which binary cross-entropy loss is employed instead of the widely used InfoNCE loss, to provide independent loss contributions from different prototypes, which is important in real scenarios in which multiple labels may apply to unsupervised data frames.
A final stage of fine-tuning with just a small amount of labeled data yields a very high performing SED model.
On like-for-like tests using the DESED task, our method achieves a PSDS1 score of 62.5\%, surpassing current state-of-the-art models and demonstrating the superiority of the proposed technique.
\end{abstract}


\section{Introduction}
Sound is ever-present in our daily lives, and sound event detection~(SED) aims to identify the onset and offset of sound events within audio signals.
In recent years, mainstream SED algorithms have been based on deep learning  strategies~\cite{cakir2017convolutional,sed-tutorial}, which typically rely on manually annotated sound events, either at the frame level (strong labels) or clip level (weak labels).
However, the cost of annotating sound events, especially strong label annotation, is particularly high~\cite{audioset-strong}.
Moreover, since the sound classes vary across different scenarios and applications, there are no universally applicable SED datasets. 
Each application requires specific data collection to meet its specific requirements~\cite{sed-tutorial}.
These factors result in a limited amount of labeled data for SED task in specific scenarios, significantly hindering the application of SED.
By contrast, collecting a sufficiently large volume of unlabeled data for a specific scenario is often more feasible. 
Thus, a major challenge for SED is how to leverage large amounts of unlabeled data, and small amounts of labeled data effectively to train SED models.

The mainstream approaches to utilize unlabeled data in SED task involve semi-supervised algorithms such as mean-teacher~\cite{tarvainen2017mean}, interpolation consistency training (ICT)~\cite{verma2022interpolation,ict}, and self-training~\cite{Ebbers2022}, but the effectiveness of the semi-supervised algorithms is constrained by the quantity and quality of labeled data.
On the other hand, self-supervised paradigms have recently gained widespread application in audio tasks, inspired by their success in computer vision~\cite{he2022masked}, natural language processing~(NLP)~\cite{DevlinCLT19}, and automatic speech recognition~(ASR)~\cite{baevski2020wav2vec,hubert,wavlm}.
These approaches learn general data representations from unlabeled examples, then fine-tune the model on labeled data.
Some of these methods adopt a contrastive learning framework~\cite{cola, 9415009, byola}, while others are based on masked audio models~\cite{gong2022ssast,baade22_interspeech,pmlr-v202-chen23ag}, predicting the content of the masked region, similar to masked language models~\cite{DevlinCLT19} in NLP. 
However, most of these works are evaluated on clip-level audio tasks, whereas SED requires a fine-grained, frame-level representations. 
Moreover, the polyphonic nature of SED task, where multiple sound events may occur simultaneously, increases the complexity of applying self-supervised strategies.
One innovative approach to apply a self-supervised strategy to SED is MAT-SED~\cite{mat-sed}, which employs a masked reconstruction task to enhance the temporal modeling capability and generalization of a Transformer-based context network.
Despite good performance, MAT-SED requires a fixed pre-trained encoder during self-supervised learning, ruling out end-to-end training.
Additionally, the reconstruction task may excessively focus on the accuracy of low-level features, failing to effectively learn high-level audio semantic abstractions~\cite{pmlr-v202-chen23ag}.

In this work, we propose a Prototype based Masked Audio Model~(PMAM) algorithm for self-supervised representation learning of SED. 
Specially, we utilize a Gaussian Mixture Model~(GMM) to model prototypical distributions of latent embeddings, thereby providing semantically rich frame-level pseudo labels for masked audio model training.
The prototypical distribution modeling and the masked audio model training are performed iteratively to enhance the quality of pseudo labels.
Additionally, we employ a prototype-wise binary cross-entropy loss to enable independent predictions of pseudo labels for different prototypes.
Experimental results show that PMAM achieves a PSDS1 score of 62.5\% on the DESED dataset, surpassing previous state-of-the-art methods.

\section{Model}
\begin{figure*}[t]
  \centering
  \includegraphics[width=0.63\linewidth]{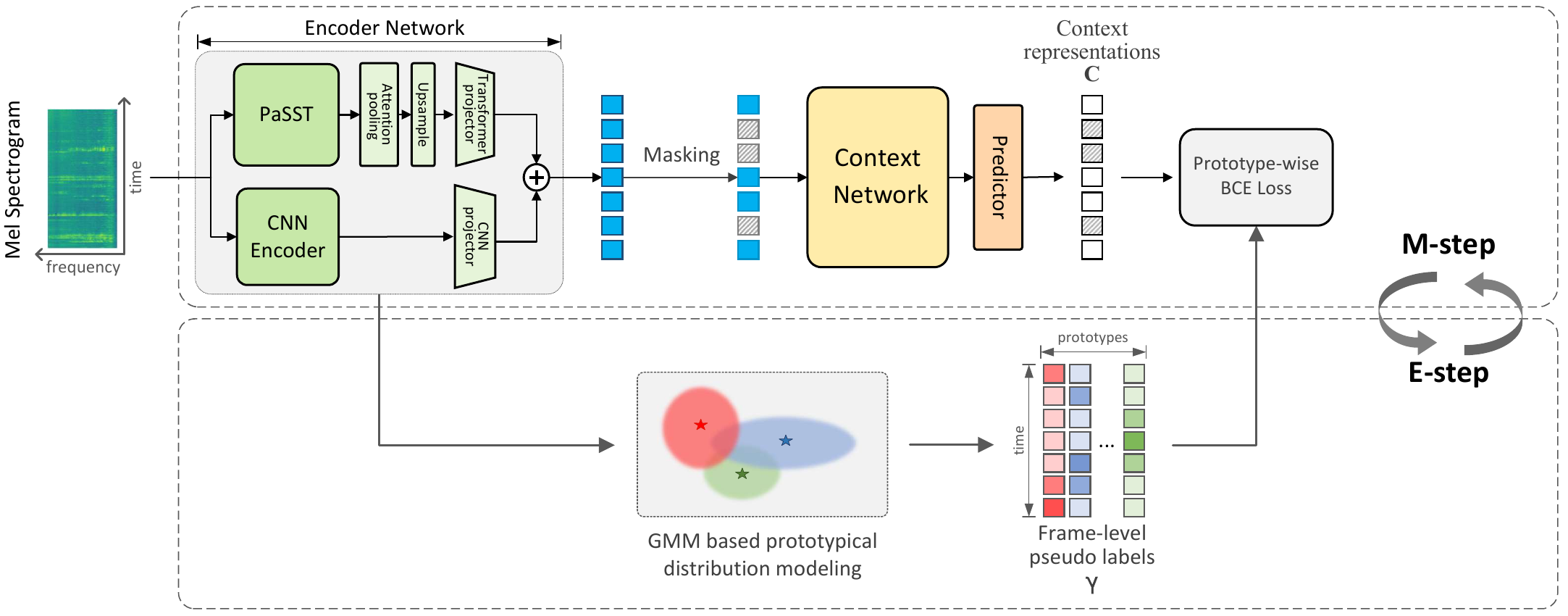}
  \caption{The proposed self-supervised iterative PMAM framework. 
  The E-step (bottom) extracts frame-level pseudo labels from latent embeddings using the prototypical distribution modeling module, to train the masked audio model, which predicts pseudo labels of the masked frames during the M-step (top).
  }
  \label{structure}
\end{figure*}

The model of PMAM is mainly composed of an encoder network and a context network, as illustrated in Fig~\ref{structure}.
The encoder network is responsible for extracting frame-level latent embeddings from the spectrogram, whereas the context network models the temporal dependencies of sound events using the masked audio model task.

PMAM requires the encoder to transform spectrograms into token sequences along the time dimension.
We implement a dual-branch encoder structure, with a Transformer branch and a CNN branch in parallel, which is  a prevalent structure in recent research~\cite{ATST, Schmid2024, Cornell2024}.
For the Transformer branch, we employ the PaSST~\cite{koutini2021efficient} architecture, which is based on the Audio Spectrogram Transformer (AST)~\cite{gong2021astaudiospectrogramtransformer}, and 
the PaSST model is initialized with pre-trained weights from AudioSet~\cite{gemmeke2017audio}.
The output of PaSST is token sequences with both time and frequency dimensions.
Then attention-based pooling is used to aggregate tokens across the frequency dimension.
Specifically, a multi-head attention module is employed, where tokens at the same temporal position but different frequencies act as keys and values, and a learnable embedding as the query. 
Subsequently, linear interpolation-based upsampling along time dimension is used to enhance features' time resolution.
Features from the CNN and Transformer branches are projected into a uniform feature space through linear projection and then merged through summation.

Latent embeddings are masked along the time dimension~\cite{mat-sed} before being fed into the context network. In the context network, a Transformer with relative position encoding~\cite{dai2019transformer} is used for modelling the temporal dependencies.
Following the context network, a linear layer serves as the predictor, outputting the context representations $\mathbf{C}=[\mathbf{c}_1, \mathbf{c}_2, ..., \mathbf{c}_T ]$ for self-supervised loss computation.

\section{Training}
We employ the frame-wise masked audio model task for self-supervised training, requiring the model to predict the content of masked frames, thereby enhancing its ability to capture temporal dependencies within sound events.
Unlike reconstruction based masked audio models, which predict either the masked spectrogram patches~\cite{gong2022ssast} or masked latent embeddings~\cite{mat-sed}, we utilize the prototype based approach, requiring the prediction of prototypes corresponding to the masked frames.
In the latent space, embeddings for a group of semantically similar frames tend to be closer, and the center of a group is termed a prototype in previous works~\cite{li2021prototypical}. 
Given the rich semantic information of prototypes, employing them as pseudo labels is eminently logical.

Prototype based masked audio models have been widely applied in ASR~\cite{hubert,wavlm}. 
These methods typically utilize clustering algorithms, usually K-means, to obtain prototypes and employ the InfoNCE loss~\cite{infoNCE} to bring an embedding closer to its corresponding prototype while separating it from other prototypes.
Despite their success in ASR, adapting these strategies to SED task introduces unique challenges. 
For ASR, each frame corresponds to a specific phoneme. 
However, for SED task, the polyphonic nature of sound events, i.e. multiple events can occur simultaneously, means a single frame might relate to multiple prototypes, diverging fundamentally from prior uses of prototype based strategies. 
The proposed PMAM algorithm is specifically designed to adapt the prototype-based approach to the complexities of polyphonic sound events, detailed as follows.

\subsection{Prototypical Distribution Modeling} \label{section:GMM}
Since each sound event corresponds to at least one prototype, a single frame may correspond to multiple prototypes when sound events overlap, a scenario that typical clustering algorithms like K-means cannot handle.
To address this issue, we represent a prototype not merely as an embedding but as a distribution of the corresponding group in the latent space. 
Specifically, let $\mathbb{Z}$ denote the latent space spanned by the frame-level embeddings from the encoder network, and $\theta_k$ denotes the k-th prototype in the total K prototypes.
The probability density of a sample $\mathbf{z} \in \mathbb{Z}$  is given by
    \begin{equation}
    p({\mathbf{z}}) = \sum\limits_{k = 1}^K {p({\theta _k})p(\mathbf{z}|{\theta _k})} \label{pz}
    \end{equation}
where $p({\theta _k})$ is the prior probability of prototype $\theta _k$, and $p(\mathbf{z}|\theta_k)$ is the corresponding conditional distribution.
The proportion of prototype $k$ in feature $\mathbf{z}$, denoted as ${\gamma_k}(\mathbf{z})$, is calculated as follows:
    \begin{equation}
        {\gamma _k}({\bf{z}}) = \frac{{p({\theta _k})p(\mathbf{z}|{\theta _k})}}{{\sum\limits_{l = 1}^K {p({\theta _l})p(\mathbf{z}|{\theta _l})} }} \label{pseudo-label}
    \end{equation}
 In this work, we utilize $\bm{\gamma}(\mathbf{z}) = [{\gamma _1}(\mathbf{z}), {\gamma _2}(\mathbf{z}), ...,{\gamma _K}(\mathbf{z})] $ as the pseudo labels for feature $\mathbf{z}$.

Furthermore, we assume that prototypical conditional distribution $p(\mathbf{z}|{\theta _k})$ follows a Gaussian distribution $\mathcal{N}(\left. \mathbf{z} \right|{\bm{\mu} _k},{\mathbf{\Sigma} _k})$ with mean $\bm{\mu}_k$ and covariance $\mathbf{\Sigma}_k$, then \eqref{pz} is equivalent to a Gaussian Mixture Model. 
A sound event is thus represented by one or more corresponding Gaussian distributions, and the mixture of various Gaussian distributions effectively models scenarios with multiple overlapping sound events.

\subsection{Objective}
InfoNCE loss, typically used in previous works of ASR, requires embeddings that relate to one positive prototype and several negative prototypes.
For single sounds it works well, but where multiple sound events can occur simultaneously, more than one positive prototype exists, hence InfoNCE loss cannot be used. 
One method to address this is to just utilize the similarity between a embedding and each separate prototype to independently predict pseudo labels of specific prototypes, analogous to multi-label classification.
To support independent predictions for different prototypes, we propose a prototype-wise binary cross-entropy~(BCE) loss function as follows
    \begin{align}
        p_{t,k} &= \sigma([2 \cdot r(sim(\mathbf{c}_t,\bm{\mu} _k)) - 1]/\tau ) \label{loss1}\\
        L_{t,k} &= -\gamma _{t,k}\log(p_{t,k}) - (1 - \gamma _{t,k}) \log( 1 - p_{t,k}) \label{loss2}
    \end{align}
where $sim(\cdot, \cdot)$ computes the cosine similarity, $r(\cdot)$ is the leaky ReLU activation function, $\tau = 0.1$ scales the logit, and $\sigma(\cdot)$ is the sigmoid function.
Equation \eqref{loss1} provides the model’s prediction for prototype k based on the cosine similarity between $\mathbf{c}_t$ and $\bm{\mu} _k$: when they are orthogonal, the cosine similarity is 0, and $p_{t,k}$ approaches 0; when aligned, $p_{t,k}$ approaches 1.
Equation \eqref{loss2} applies binary cross-entropy to compute the loss between $p_{t,k}$ and the pseudo-label $\gamma _{t,k}$. The calculation of $L_{t,k}$ focuses solely on the similarity between $\mathbf{c}_t$ and $\bm{\mu} _k$,  thereby making it apt for multi-label SED tasks.

The overall loss function over the total dataset $\mathcal{D}$, the set of masked frame indices $M_x$, and all prototypes $k$ is: 
\begin{equation}
    L = \sum\limits_{x \in D} {\sum\limits_{t \in {M_x}} {\sum\limits_k {{L_{tk}}} } } 
\end{equation}
Effective prediction of pseudo labels for masked frames requires not only good representations from the encoder network but also the context network’s capability to model temporal dependencies, thereby enhancing the overall performance of the model.

\subsection{Iterative Update of Pseudo Labels}
The quality of the pseudo labels depend on latent embeddings generated by the encoder network. 
In the first iteration, the output features from PaSST are average pooled, upsampled and then used to generate initial pseudo labels via prototypical distribution modeling. 
Upon completing the masked audio model training of the first iteration, the pseudo labels are regenerated using representations extracted by the updated encoder network.
The prototypical distribution modeling and the masked audio model training are performed iteratively, similar to E-step and M-step in the expectation-maximization (EM) algorithm~\cite{moon1996expectation}, to enhance the quality of pseudo labels.

\subsection{Semi-supervised Fine-tuning}
After the self-supervised training stage, fine-tuning the model with only a small amount of labeled data suffices to achieve a well-performing SED model.
During the semi-supervised fine-tuning stage, the predictor on the top of the self-supervised model is replaced with a classifier composed of a fully connected layer and a sigmoid activation for event prediction.
We employ the widely used mean-teacher model~\cite{tarvainen2017mean} for semi-supervised learning in this stage.

\section{Experiments}
\subsection{Experiment Setting}
Experiments are conducted on the DESED dataset~\cite{Turpault2019_DCASE}, designed for  sound event detection in domestic environments.
The training set consists of  1578 weakly-labeled clips, 3470 strongly-labeled clips, 14412 unlabeled in-domain clips, and 10000 synthetic clips generated from 1011 audio files. 
Model performance is evaluated on a validation set with 1168 strongly-labeled clips. 
We assess the model performance using the PSDS1 score~\cite{psds} and apply two post-processing methods: the classical median filter and the novel Sound Event Bounding Boxes (SEBB) method based on change point detection~\cite{ebbers2024soundeventboundingboxes}.

The feature extraction follows the setting of \cite{koutini2021efficient}.
The CNN branch in the encoder follows the DCASE2024 task 4 baseline~\cite{Cornell2024}, including 7 convolutional layers. The context network  consists of 3 Transformer blocks. 
The GMM module has 30 Gaussian components and is trained using the EM algorithm~\cite{moon1996expectation}.
We use the AdamW optimizer, with a learning rate of $1\times 10^{-5}$ for the PaSST module and $2\times 10^{-4}$ for the rest. 
The batch size is set to 18. 

The self-supervised training takes two iterations, and each iteration lasts for 30 epochs. In this stage, we fine-tune the last 4 transformer blocks of PaSST using low-rank adaptation~(LoRA)~\cite{hu2022lora} with rank~=~8, which trains only additional low-rank matrices while keeping the original pre-trained parameters fixed, significantly improving training efficiency.
Masking during the masked audio model training involved a 75\% ratio with block-wise mask strategy with size 10~\cite{mat-sed}. The fine-tuning stage lasts 45 epochs, with the first 15 epochs training only the top linear classifier.

\subsection{Comparing with the SOTA Models}
Table~\ref{tab1} compares the performance between PMAM and previous SOTA models in the validation set.
 Models fine-tuned with first and second iteration weights are denoted as PMAM\textsubscript{iter1} and PMAM\textsubscript{iter2}, respectively; PMAM\textsubscript{iter0} denotes fine-tuned results without self-supervised training.
As shown in the table, the performance of PMAM\textsubscript{iter0} is far more behind than other SOTA models.
However, after the first round of self-supervised training, PMAM\textsubscript{iter1} surpasses previous SOTA models with both post-processing methods, demonstrating the effectiveness of the proposed self-supervised algorithm.
After the second iteration, PMAM\textsubscript{iter2} shows a slight improvement over PMAM\textsubscript{iter1}, benefiting from the enhanced quality of pseudo labels. Notably, we only used the basic mean-teacher algorithm for simplicity during semi-supervised fine-tuning.
Incorporating other semi-supervised  strategies, such as ICT, might yield additional improvements.

\begin{table}[t]
    \footnotesize
    \centering
    \caption{Comparison with state-of-the-art single SED systems}
    \renewcommand\arraystretch{1}
    \begin{tabular}{c  c | c c}
    \Xhline{1pt}
        \multirow{2}{*}{\textbf{Model}} &  \multirow{2}{*}{\textbf{Self-supervised}} & \multicolumn{2}{c}{\textbf{PSDS1}} \\
        \cline{3-4}
        && \makecell[c]{median filter} & \makecell[c]{SEBB} \\
    \hline
        System of CP-JKU~\cite{Schmid2024} &  \XSolidBrush   & 54.8\% & 61.7\%\\
        ATST-SED~\cite{ATST} & \XSolidBrush    & 58.3\% & -\\
        MAT-SED~\cite{mat-sed} & \Checkmark   &  58.7\%   & 60.2\% \\
    \hline
    PMAM\textsubscript{iter0}  & \XSolidBrush   & 56.3\%   & 58.5\%  \\
    PMAM\textsubscript{iter1} & \Checkmark   &  59.4\%  &  62.2\% \\
    PMAM\textsubscript{iter2} & \Checkmark   & \textbf{59.7\%} & \textbf{62.5\%} \\
    \Xhline{1pt}
    \end{tabular}
    \label{tab1}
\end{table}

\begin{table}[t]
    \footnotesize
    \centering
    \caption{Ablation study of the proposed PMAM algorithm. Median filter is used for post-processing.}
    \renewcommand\arraystretch{1}
    \begin{tabular}{c c c c}
    \Xhline{1pt}
        \textbf{\makecell[c]{Masked Audio\\ Model}} & \textbf{\makecell[c]{Prototype  \\Modelling}} & \textbf{\makecell[c]{Loss \\ Function}} & \textbf{PSDS1} \\
    \hline
        \XSolidBrush & GMM & prototype-wise BCE & 57.5\%\\
        \Checkmark & K-Means & prototype-wise BCE & 58.9\%\\
        \Checkmark & GMM & InfoNCE & 59.2\%\\
    \hline
        \Checkmark & GMM & prototype-wise BCE & \textbf{59.7\%}\\
    \Xhline{1pt}
    \end{tabular}
    \label{tab:ablations}
\end{table}


\subsection{Ablations}
To assess the contributions of different components of our method, we conducted ablation studies, as shown in Table~\ref{tab:ablations}. 

\subsubsection{Masked audio model}
The masked audio model forms the cornerstone of our algorithm, which guides the model to learn the temporal dependencies of sound events.
As shown in Table~\ref{tab:ablations}, the model without masking achieves only 57.5\% PSDS1, significantly lower than PMAM\textsubscript{iter2}, which shows the masked audio model's critical role in our method. 
Notably, even the performance without masking substantially exceeds that of PMAM\textsubscript{iter0}, suggesting that prototype based pseudo labels can enhance model performance beyond just semi-supervised fine-tuning, even without the masked audio model.

\subsubsection{Prototype modelling algorithm}
Given the polyphonic nature of sound events, the GMM based prototypical distribution modeling proves more effective than clustering algorithms like K-means for generating pseudo labels, as discussed in section \ref{section:GMM}.
We experiment by substituting GMM with K-means for pseudo-label generation, leading to a decrement of 0.8\% compared to using GMM based prototypical distribution.

\subsubsection{Loss function}
A comparison of loss functions during self-supervised training is also presented in Table~\ref{tab:ablations}. 
The results indicate that our proposed prototype-wise BCE loss improves performance by 0.5\% over the InfoNCE loss, confirming its suitability for SED tasks.

\begin{figure}[t]
  \centering
  \includegraphics[width=0.7\linewidth]{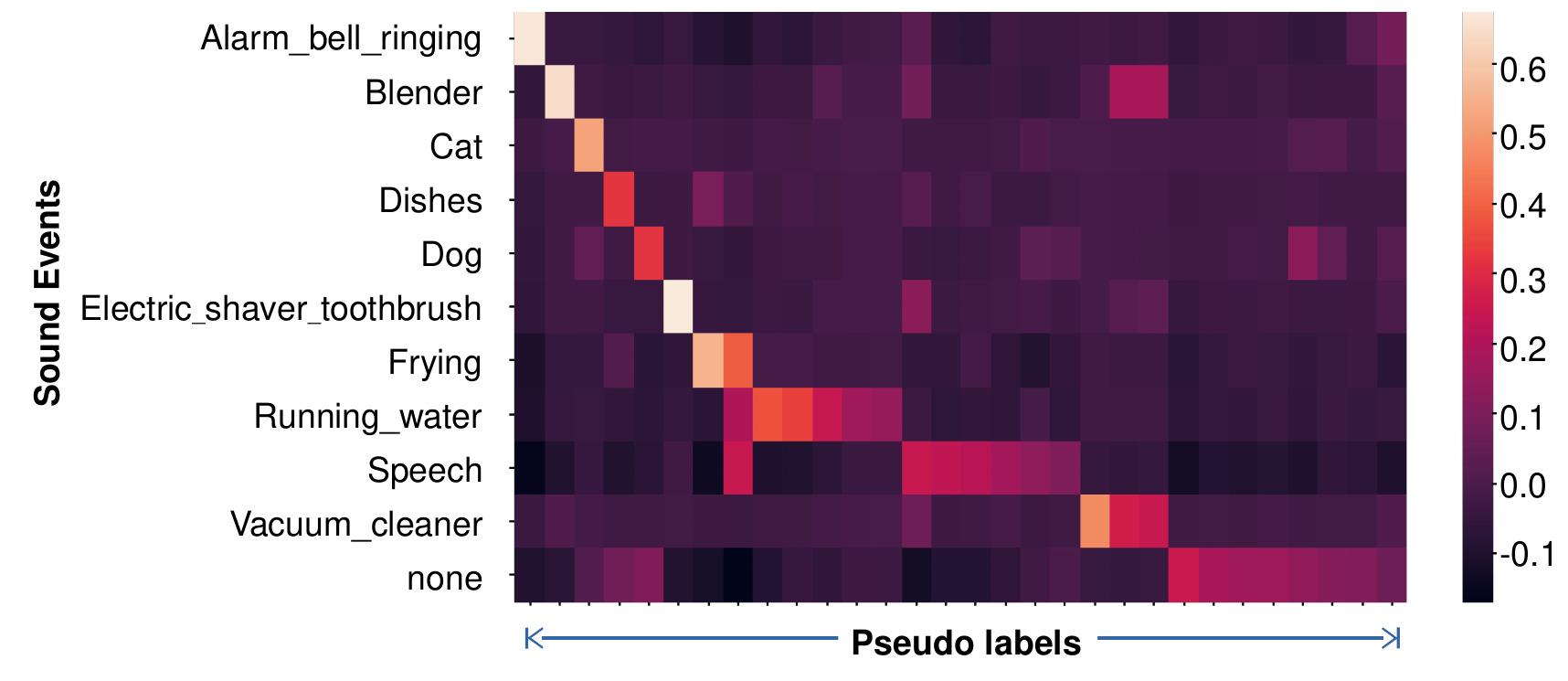}
  \vspace{-2mm}
  \caption{The point-biserial correlation coefficient matrix between prototype based pseudo labels in the second iteration and real labels. `None' represents the label of frames when no event occurred. The pseudo labels are reordered to match the sequence of the real labels for better revealing the correlation.}
  \label{corr_matrix}
\end{figure}
\begin{figure}[t]
  \centering
  \includegraphics[width=\linewidth]{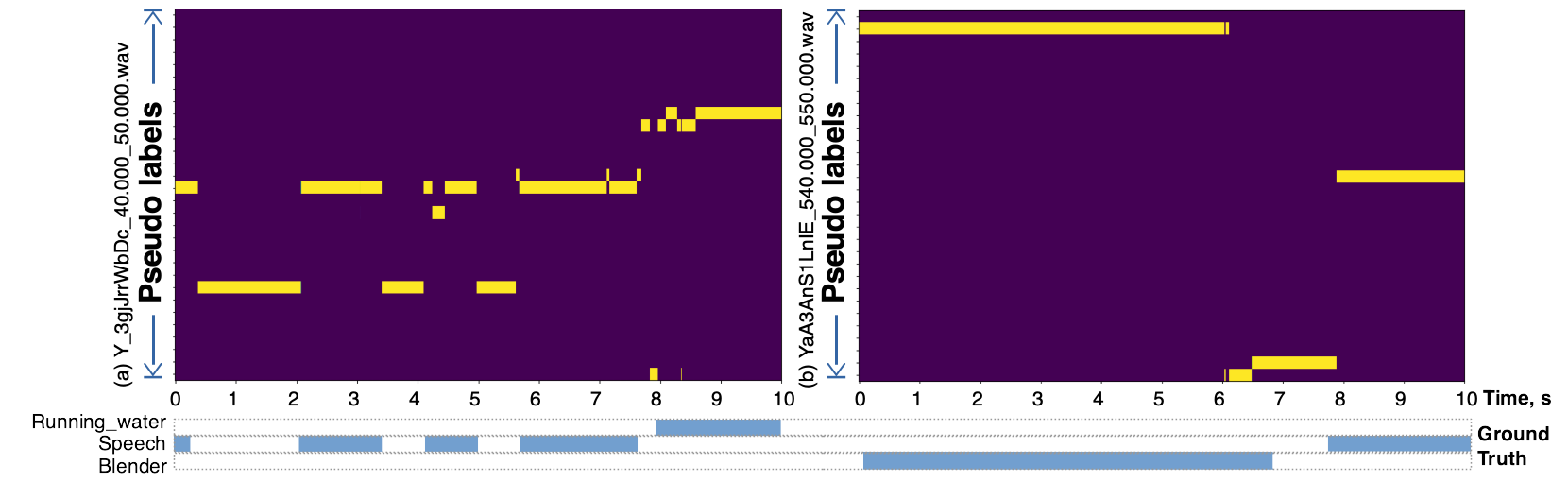}
  \caption{Pseudo labels and ground truth corresponding to audio samples in the second iteration. }
  \vspace{-2mm}
  \label{timestamp}
\end{figure}

\subsection{Analysis and Visualization of Pseudo Labels}
We analyzed the relationship between prototype based pseudo labels and real labels using the point-biserial correlation coefficient, as depicted in Fig~\ref{corr_matrix}.
Most prototypes show strong correlations with specific sound events, indicating the their rich semantic information.
Additionally, we can see from Fig~\ref{corr_matrix} that pseudo labels provide insights beyond the real labels. 
For instance, sound events ‘Running\_water’ and ‘Speech’ are correlated with multiple prototypes, suggesting high intra-class variance of these sound events.
Furthermore, some prototypes associated with ‘Vacuum\_cleaner’ also show clear correlation with ‘Blender’, due to the similar acoustic characteristics between two sound events. 
This demonstrates that pseudo labels can also capture inter-class similarity among different sound events.

Fig~\ref{timestamp} displays the pseudo labels and ground truth of randomly selected audio samples. We can see from Fig~\ref{timestamp} that most sound event timestamps align closely with those of the pseudo labels, with similar start and end times.
We  can also observe that the prototypes activated during specific events correlate strongly with those events. 
These observations indicate that prototype based pseudo labels contain both localization and rich semantic information, which is crucial for SED task.

\section{Conclusion}
In this paper, we introduce PMAM, a framework for self-supervised representation learning in sound event detection. 
By leveraging the prototypical distribution modeling, we extract semantically rich pseudo labels, which are then used to train the masked audio model. 
The use of prototype-wise BCE loss enables independent loss computation of different prototypes.
Experimental results demonstrate that PMAM significantly enhances the performance of SED models compared to solely semi-supervised fine-tuning, revealing the substantial potential of unsupervised algorithms for SED task, especially in scenarios with limited labeled data.
Moving forward, we plan to further explore the application of unsupervised algorithms for audio tasks.
\bibliographystyle{IEEEtran}
\bibliography{wpref}

\end{document}